# RQA Application for the Monitoring of Financial and Commodity markets state

Sergii Piskun[1*], Oleksandr Piskun[1], Dmitry Chabanenko[2]

[1]Cherkasy Banking Institute of Banking University of the National Bank of Ukraine, Cherkasy, Ukraine
[2]Cherkasy National University named after B. Khmelnytsky, Cherkasy, Ukraine

Nowadays, when crashes and crises are rather frequent events, an effective monitoring system for the international financial market is needed. Modern nonlinear methods, such as Recurrence Quantification Analysis (RQA), demonstrate the ability to reveal the regularities of the system behavior. Thus, they can be useful for the analysis of the market state in real time. In present paper we did an effort to apply the RQA for the purpose of economic time series monitoring. 12 stock indexes, 6 currency pairs and 4 commodities were taken for the study.

## 1. Introduction

Modern economic system, especially it's financial sector, is much more complex than it was some decades ago. Now, every person is direct or indirect investor. The majority of people put their money to the bank in order to have an interest. Bank provides loans for the enterprises. They expand their business and issue stocks or bonds. So, every link in this chain depends on the situation on the financial market. Some people prefer to have their savings just in a bank cell. This case doesn't prevent the market risk either. The value of such capital is determined by the currency rate, gold price etc. Those, who have certificates of the special fund, depend on the average-weighted price of the basket of financial assets that the fund contains. Moreover, insurance-holders are also under the risk of market changing. Insurance company invests collected money in gold, currency, stocks, bank deposits and real estate. So, people's financial safety depends on the market state as well. Such situation is regular not only for the developed countries. With the help of internet, international corporations, liberalization of the protectionism laws the majority of countries are involved to the process of forming of the one world financial market. That it why, the problem of adequate monitoring of the market state appears to be rather urgent nowadays.

Classic econometric methods can hardly describe financial market effectively, because they were developed for much more simpler markets with the lower level of globalization, involvement, regulation, eventually complexity. Here we need nonlinear methods, that don't require simplifications of the data analysis, such as Recurrence Plot (RP) and Recurrence Quantification Analysis (RQA). They came from physics and were widely used in biology, physiology, ecology, earth sciences [e.g., Zbilut *et al.*, 2004; Trauth *et al.*, 2003; Marwan *et al.*, 2007; Schinkel *et al.*, 2007; Facchini *et al.*, 2007] and in economics as well. Holyst *et al.* [2001], Gilmore [1996] and McKenzie [2001] used these techniques for chaos testing in financial time series. Antoniou and Vorlow [2000], Gilmore [2001], Kyrtsou *et al.* [2009] and Belaire-Franch *et al.* [2002, 2004] applied them for the examining of nonlinear dependencies. Kyrtsou and Vorlow [2005] discovered non-linear determinism and complex dynamics of US macroeconomic time series. Holyst and Zebrowska [2000], Pecar [2004] and Zbilut [2005] exploited methods for revealing of nature of financial market behavior. Strozzi *et al.* [2002, 2007] used RQA for the measurement of the financial data volatility and correlation between currency time series. Fabretti and Ausloos [2005] applied RP and RQA for the critical regime detection in financial markets and estimation of the bubble initial time. Bastos and Caiado [2011] studied interdependencies between stock markets and their behavior during critical events.

---

[*]*Corresponding author.* E-mail address: sergii_com@ukr.net

The aim of the present work is to study the ability of laminarity measure to reflect the market state changes in real time.

## 2. Recurrence plot and recurrence quantification analysis

Eckmann *et al.* [1987] introduced recurrence plot (RP) (fig. 1) as a tool for representation of the system behavior in m-dimensional phase-space by the 2-dimensional matrix through it's recurrences [Marwan, 2003]:

$$R_{i,j}^{m,\varepsilon_i} = \Theta(\varepsilon_i - \|\vec{x}_i - \vec{x}_j\|), \vec{x} \in \Re^m, i, j = 1,..., N \qquad (1)$$

where $N$ is a number of the considered states $x_i$, $\varepsilon_i$ is a threshold distance, $\|\cdot\|$ a norm and $\Theta(\cdot)$ the Heaviside function.

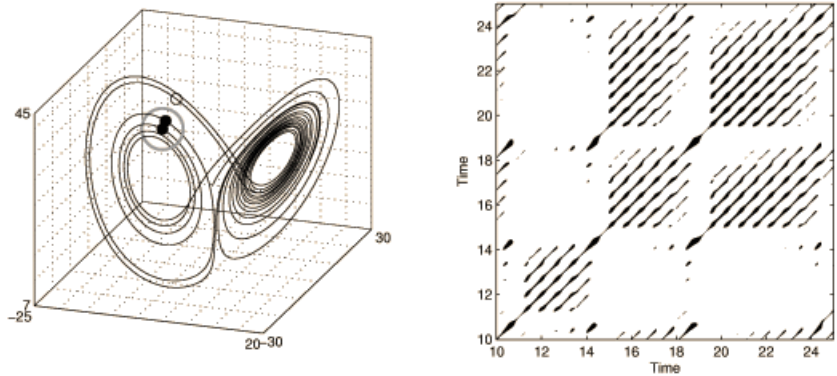

Fig. 1. The phase space trajectory of the Lorenz system and its corresponding RP [Marwan, 2003].

Zbilut and Webber [1992, 1994] developed a set of quantitative measures that are calculated in the terms of RP. They defined recurrence rate (RR), determinism (DET), average diagonal line length (L), divergence (DIV), entropy (ENTR) and trend (TREND).

RR: fraction of recurrence points and all points in the RP. It reflects the density of recurrence points.

DET: fraction of recurrence points in diagonal lines and all recurrence points. The measure indirectly shows the level of the system determinism.

L: fraction of recurrence points in diagonal lines and diagonal lines. This measure presents the average time when different segments of the system trajectory are close. So, it can serve as an average time for prediction.

DIV: inversion of the length of the longest diagonal line. It is connected with the divergence of the phase space trajectory.

ENTR: Shannon entropy of the diagonal lines distribution. It is the measure of complexity.

TREND: linear regression coefficient over the RR on each diagonal line parallel to the main diagonal (LOI). It detects the degree of the system stationarity.

Later, Gao [1999] proposed time statistics measures: Trapping time 1 (T1) and Trapping time 2 (T2).

T1: times between the recurrence points in columns. It shows the frequency of visiting of the certain point neighborhood by the system trajectory.

T2: times between the beginning points of vertical lines in columns. It eliminates the tangential motion and can be useful for the estimation of the point-wise dimension.

Marwan *et al.* [2002] worked out measures of laminarity (LAM) and trapping time (TT).

LAM: fraction of recurrence points in vertical lines and all recurrence points. It represents the level of system laminarity.

TT: faction of recurrence points in vertical lines and vertical lines. It defines the average time of the laminar periods of the system behavior.

Presented measures have formed recurrence quantification analysis (RQA).

## 3. RQA embedding parameters

One of the significant aspects of RQA applycation is the appropriate embedding parameters: dimension $m$, time-delay $\tau$ and threshold $e$. The problem of phase space reconstruction and respectively $m$ and $\tau$ determination are widely discussed in [Camastra, 2003; Kantz & Schreiber, 1997; Marwan *et al.*, 2007; Packard *et al.*, 1980; Thiel *et al.*, 2004]. The study of $e$ choice can be found in [Donner *et al.*, 2010; Gao & Jin, 2009; Koebbe & Mayer-Kress, 1992; Marwan *et al.*, 2007; Matassini *et al.*, 2002; Schinkel *et al.*, 2008; Thiel *et al.*, 2002; Zbilut &Webber 1992]. According to [Marwan, 2011] the parameters should be chosen with respect to the particular time series and to the particular problem. Moreover, the study of sensitivity of the results, obtained by means of RQA, to different embeddings is suggested.

Here, we have daily financial time series, studied by means of RQA in order to develop the real time monitoring tool. So, we should determine the appropriate embedding parameters for such kind of research. As a control time series, we take S&P500 from 09.08.1999 till 11.07.2011. First of all, let's define the time-delay $\tau$ and dimension $m$. For this, a VRA software can be used [Mhalsekar *et al.*, 2010]. Fig. 2 shows the average mutual information. We are interested in the first minimum of the function, because it occurs at the time lag that would indicate the value of $\tau$.

So, time-delay is 23. Then, we use the method of False Nearest Neighbors (FNN) to define $m$ (fig. 3).

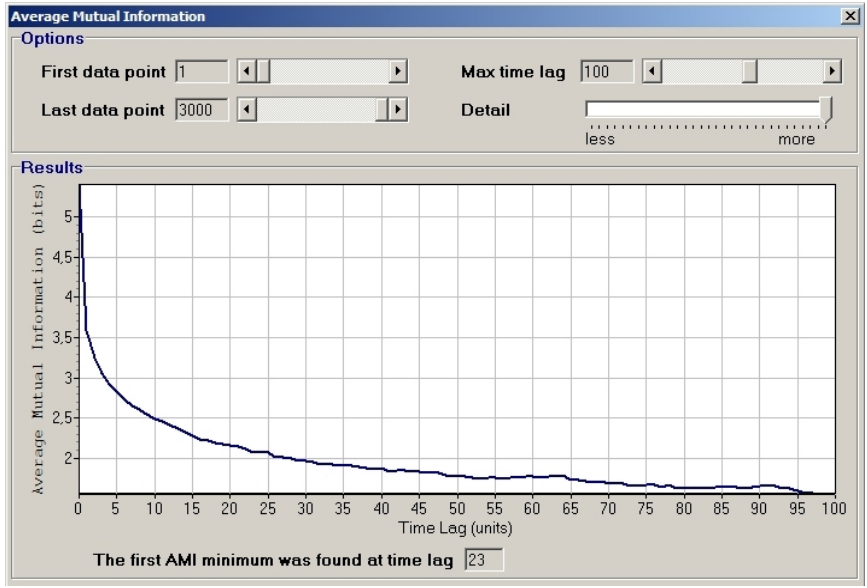

Fig. 2. Average mutual information of S&P500.

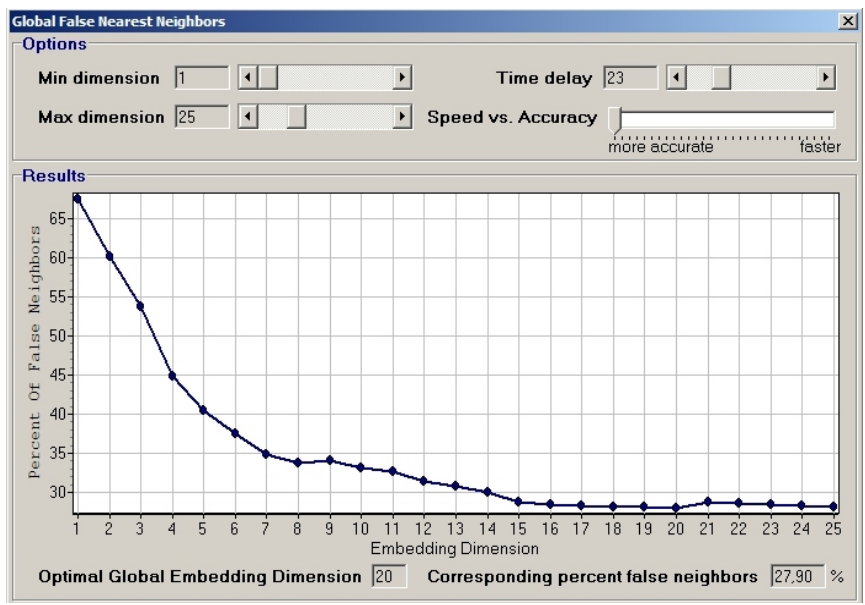

Fig. 3. Optimal global embedding dimension of S&P500 with time-delay 23.

We receive the optimal global embedding dimension equal 20. Unfortunately, with such embedding parameters we can't obtain any adequate RQA measures. Financial markets are not natural systems. Moreover, they are not only artificial ones, but contain the high percent of people's factor. Maybe because of this, embeddings, calculated with the standard methods, are not suitable for them. Nevertheless, let's try to see how sensitive LAM measure behavior is to various embedding parameters. For the etalon we take the LAM measure of S&P500 from 09.08.1999 till 11.07.2011 with dimension and delay = 1, threshold = 0,1, maximum norm. The window size *ws* was chosen 250 for the series of 1000 and more points. In such case, we eliminate the stochastic view of LAM measure and it would still react on the market changes rather quickly.

It is not necessary to check the LAM measure changes with the various time-delays, because without $m$, $\tau$ doesn't change the LAM graph. The next parameter is embedding dimension $m$. In order to determine its value we used the method of False Nearest Neighbors, realized in CRP toolbox for MATLAB® [Marwan *et al.*, 2007] (fig. 4).

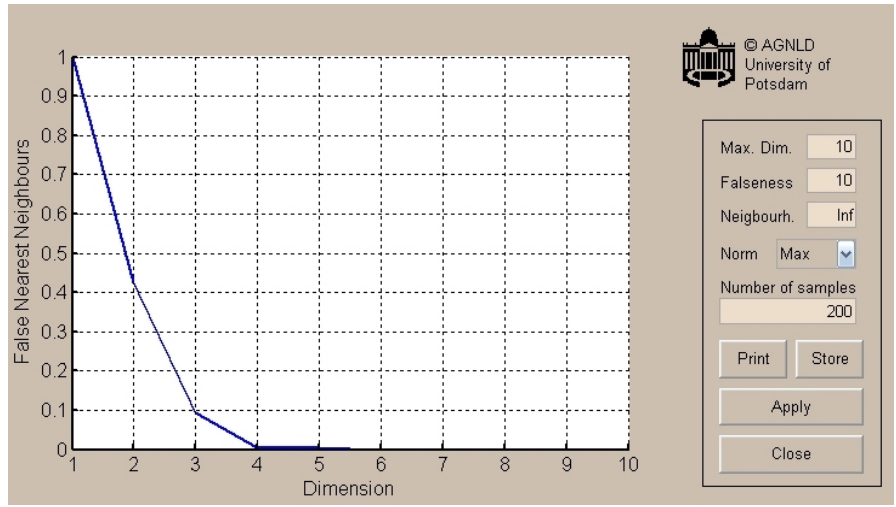

Fig. 4. Embedding dimension of S&P500.

According to FFN, the minimum value of $m$ should be 4, but lets cross-check the behavior of LAM with different $m$ (fig. 5).

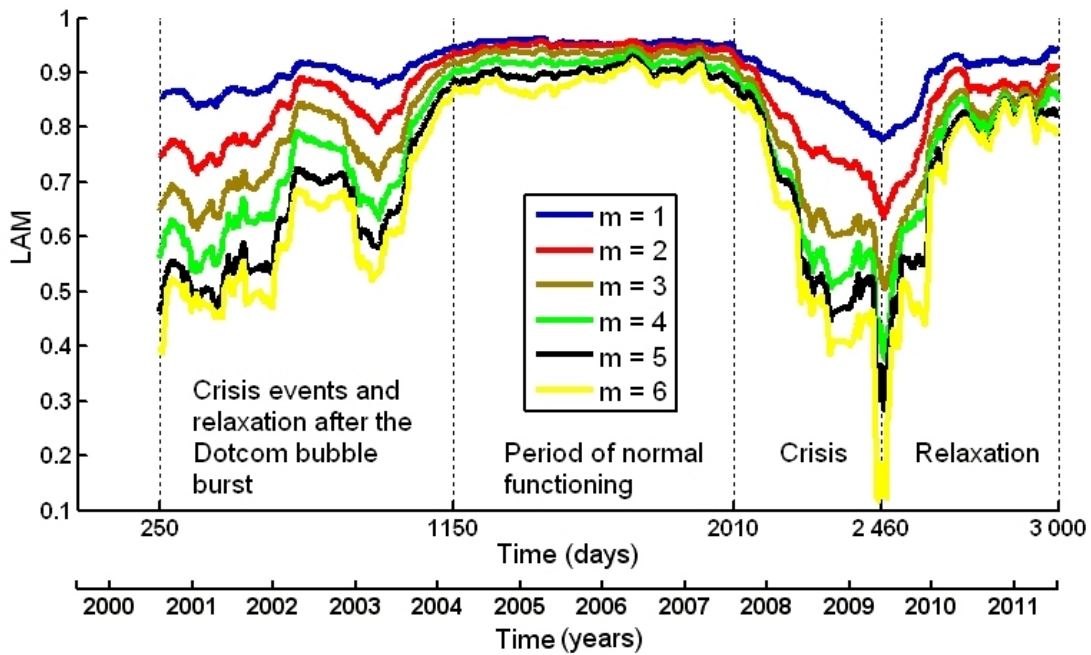

Fig. 5. LAM graphs with various embedding dimension $m$.

In [Piskun & Piskun, 2011] was shown the possibility of LAM measure to reveal different periods of financial market functioning. Fig. 5 is one more example of this feature.

Higher levels of *m* lower the LAM values and make the period from x = 2010 till 3000 hardly understandable for interpretation. When *m* = 1, we have the smooth downfall of LAM (from x = 2010 till 2460), that indicates that S&P500 experience the crisis period, and LAM increasing (from x = 2461 till 3000), that tells about the period of relaxation. Other values of *m* distort this LAM dynamics, which is important for our aim. So, despite of the FNN results, we will use parameter *m* equal 1.

Let's see how the LAM measure is changed with the various thresholds (fig. 6).

Different thresholds don't change the LAM view very much. The radius, bigger than 0,1 increases the LAM values and decreases the downfall during the crisis period (x = 2010 - 2460). The radius, smaller than 0,1 acts vice versa. It decreases the LAM values and increases the downfall. In general, we don't receive any better results with the varying the threshold values. Therefore, 0,1 is an appropriate quantity for this parameter.

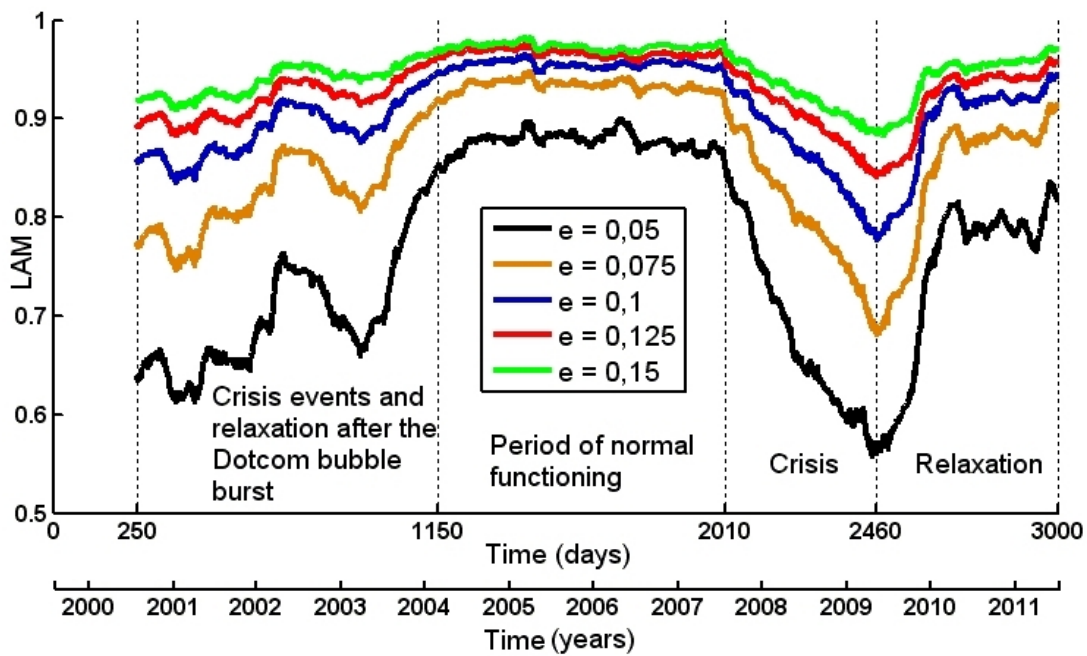

Fig. 6. LAM graphs with different thresholds *e*.

So, in the case of financial market study for the purpose of analysis and monitoring of the market state, we will use *m* = 1, *τ* = 1, *e* = 0,1 and *ws* = 250.

All further computations would be done by means of MATLAB® version of RQA developed at the University of Potsdam called CRP toolbox, which can be found at http://tocsy.agnld.uni-postdam/de [Marwan *et al.*, 2007].

Another aspect of RQA application is the appropriate RQA measures choice [Marwan, 2011]. For example, the analysis of sinus (fig. 7) requires the diagonal-based measures, because vertical lines simply don't exist. In the case of financial and commodity market study, the vertical line-based measures should be taken, as economic time series (fig. 7, 8) form only clusters of vertical or horizontal lines and no diagonal lines are presented. So, such measures as RR, LAM, TT, T1 and T2 would be suitable for the market analysis. Other measures like DET or ENTR would derive the diagonal lines from the clusters and reveal the false system properties.

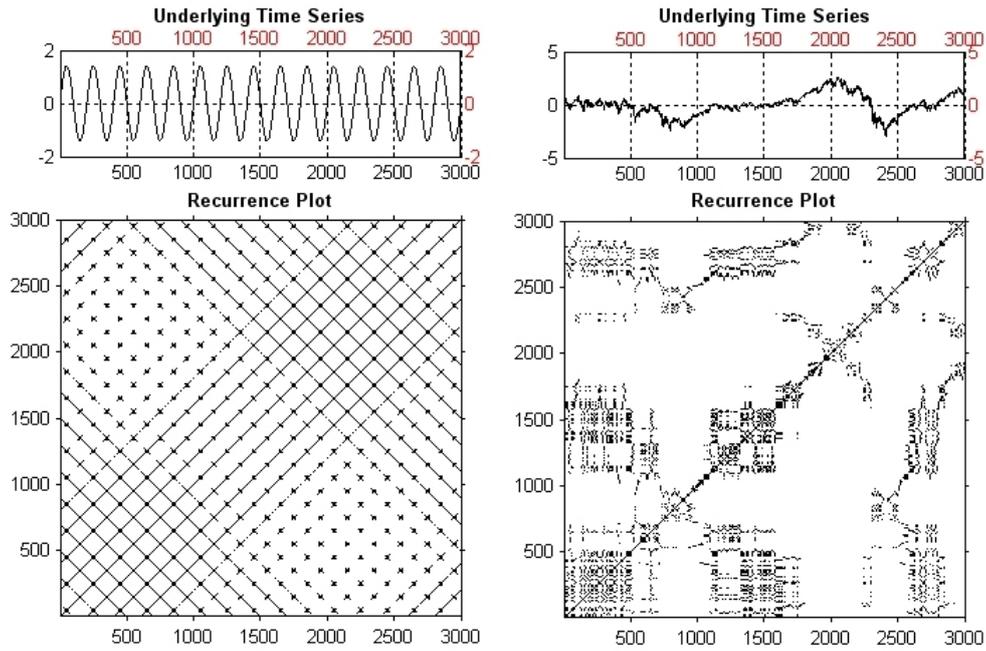

Fig. 7. Recurrence plots of sin((1:1000)*2*pi/67) and DJI from 09.08.1999 till 11.07.2011 respectively.

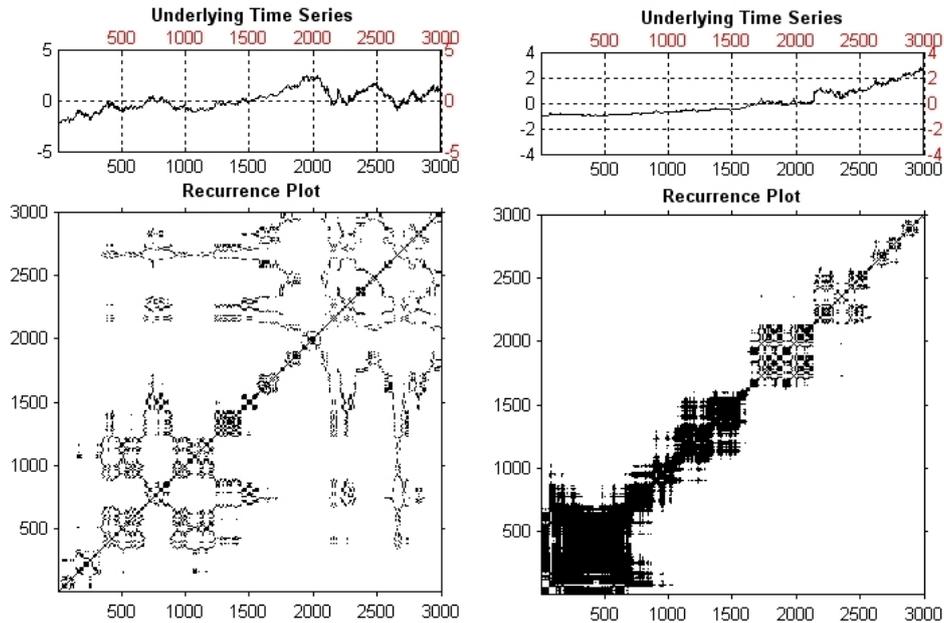

Fig. 8. Recurrence plots of eur/usd from 19.12.2002 till 11.07.2011 and Gold from 10.08.1999 till 29.07.2011 respectively.

## 4. Financial and commodity markets monitoring by means of RQA

The standard method of RQA measures calculation supposes several steps [Marwan *et al.*, 2007]. First of all, the data row is normalized. Then the RP is build. From the RP, the window (quadrate matrix), the length of which is determined by the parameter "window size", is cut off. On it basis, the calculation of measures are provided and the received points are written down. Then the window moves one point further, the measures are calculated again and the next points are written down. This process lasts to the moment of the window reaching of the last RP point. Finally we receive RQA measures. According to [Strozzi *et al.*, 2002, 2007; Fabretti & Ausloos, 2005; Piskun & Piskun, 2011] they can reveal and reflect the market state changes. For example, fig. 9 demonstrates the ability of LAM measure to distinguish different market regimes of functioning. So, RQA measures are useful for the market analysis.

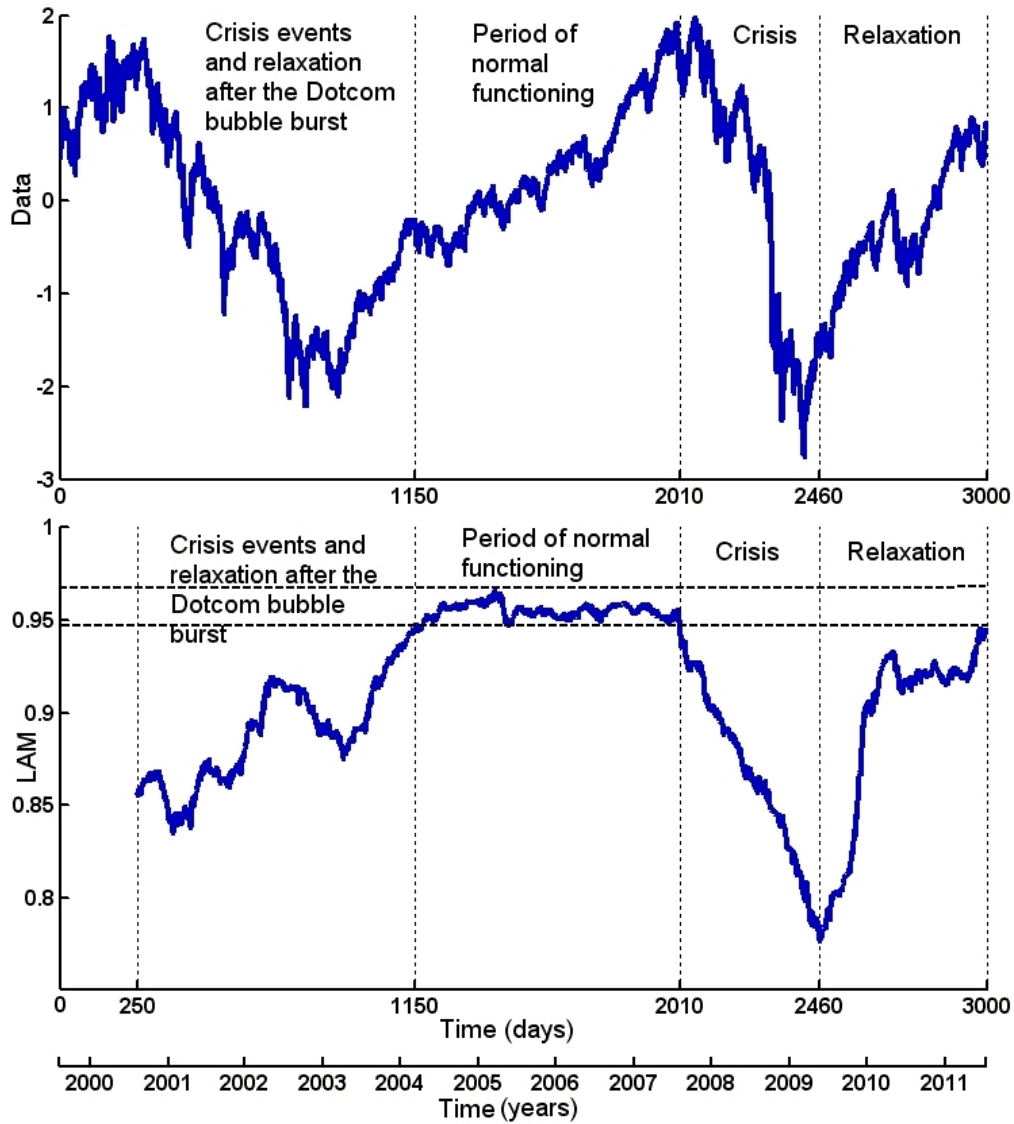

Fig. 9. S&P500 from 09.08.1999 till 11.07.2011 and corresponding LAM measure.

For the purpose of monitoring in real time, we should provide the computation of measures every day (in case of daily data). For example, we choose the row with the length of 1000 points of daily values. The last data point is today. We compute LAM measure and write down only the last point. This point reflects the current marker state, as it was calculated on the basis of the period from the present date to some time in the past. Tomorrow, the series of 1000 points moves to one point right (the next data point is added and the first one is deleted). The calculation of LAM is provided again and the last point is written down. For the future days the procedure is the same. In the result, obtained LAM graph would be a monitoring tool for the market state.

In order to study the properties of such LAM, some history of LAM points is needed. For this, we simulated the real time monitoring process. We took the series of 3000 points. Chose the length of the part of the row (LPR) that would be cut off [Piskun & Piskun, 2011] 1000 points. We took this part (from 1 to 1000), normalized it and built the PR (fig. 10).

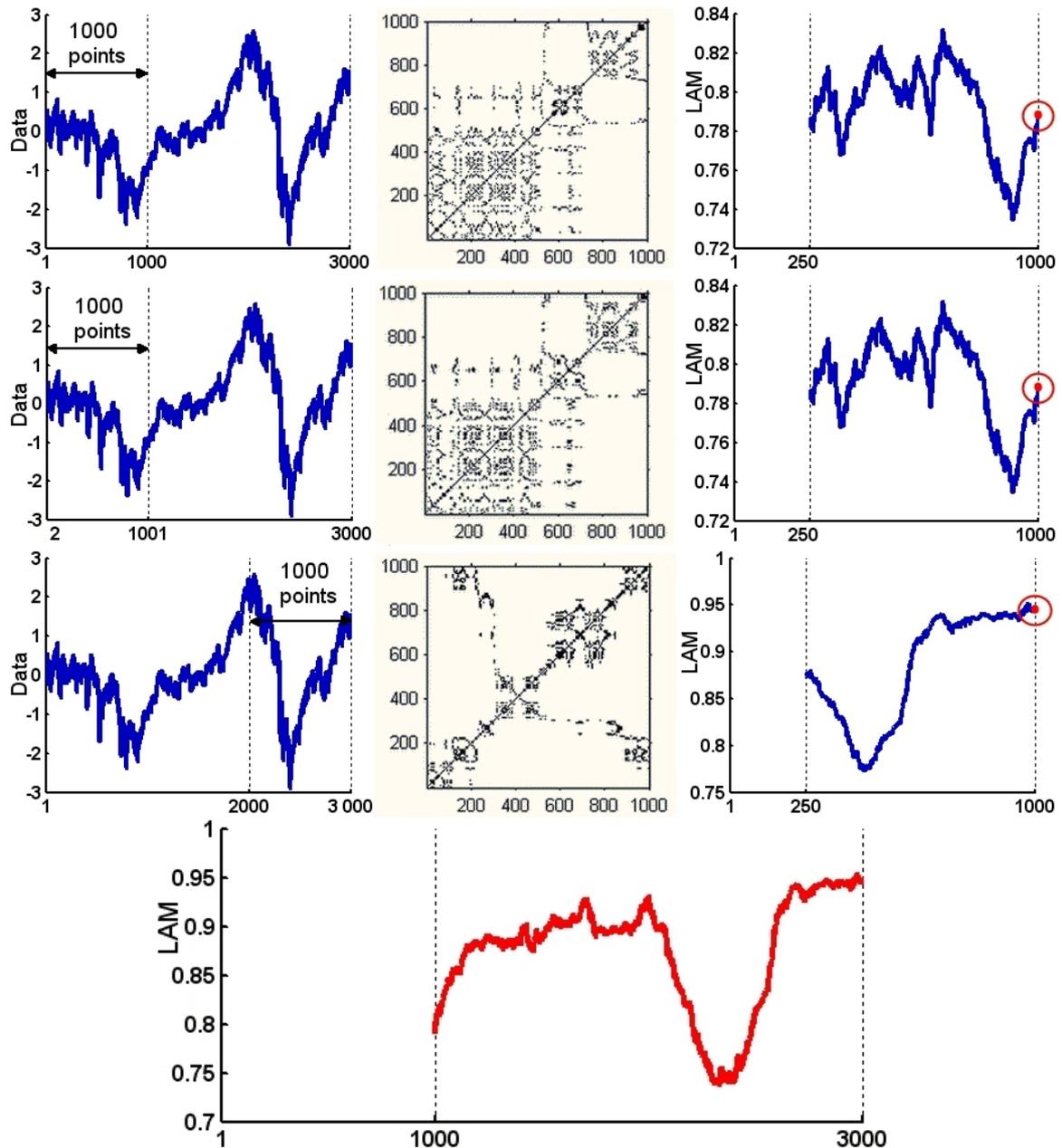

Fig. 10. The simulation of LAM measure calculation in real time on the basis of DJI from 09.08.1999 till 11.07.2011.

Then we calculated the LAM measure (with the window size = 250) and wrote down the last point. After, we took the next part of the row from 2 to 1001 points, normalized it and built the RP, calculated LAM and wrote down the last point. The finally received LAM graph (fig. 10) contains points that are calculated only on the basis of the historic date. So, we obtained the same LAM like in the case of everyday calculation in real time. The LAM graph starts from the coordinate x = 1000 in order to correspond to the data points.

Now we should check whether the new LAM has the same properties as the regular one. The simple way to do this is to compare two LAMs calculated in different ways (fig. 11). For this purpose, the row of S&P500 from 09.08.1999 till 11.07.2011 was taken. According to [Piskun & Piskun, 2011] the LPR was chosen 1500 points.

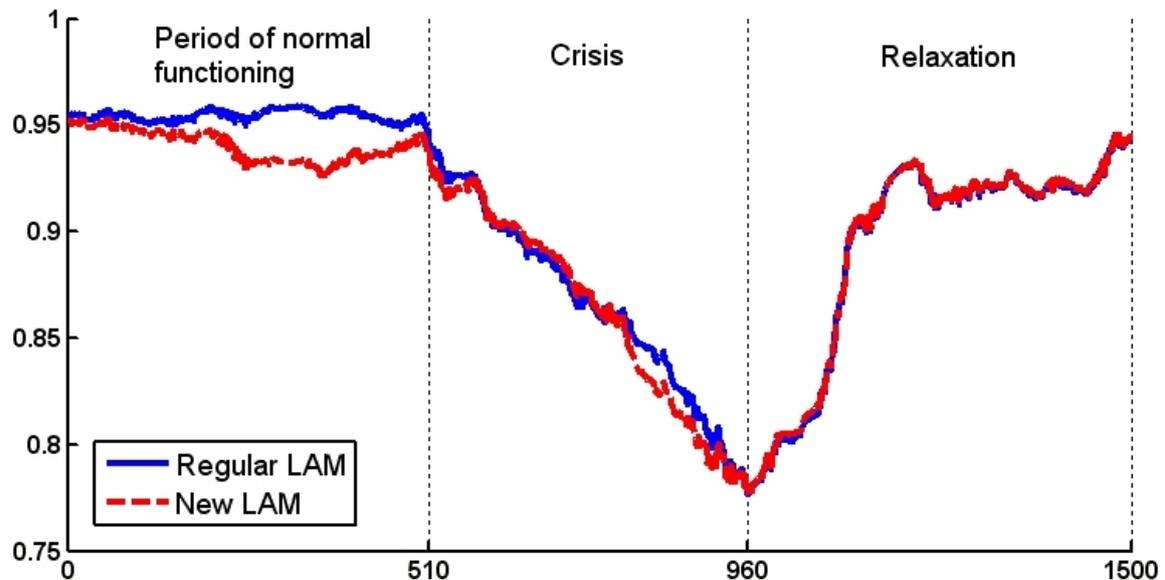

Fig. 11. S&P500 from 09.08.1999 till 11.07.2011 and corresponding LAM measures.

For the better comparison, the part of the regular LAM was cut off. Only the last 1500 points were left. As we see, two LAMs are rather similar and reveal the same regimes of market functioning. From the coordinate x = 1 till x = 510 graphs differ a little bit, during the crisis period they are more similar and from x = 960 they are approximately equal. So, the LAM measure calculated in real time is useful for the purpose of financial market monitoring.

After the study of the LAM ability to serve as a monitoring tool, let's check whether there are some quantitative patterns in LAM behavior during the different periods of market functioning. For this the series of DJI, FTSE, S&P, DAX, HIS, Nikkei, CAC, ASE, SMI, OMXC, PSI and AORD indexes were taken. Each row is 3000 points long and has the last data point on the 11.07.2011. First of all we used the regular method of RQA calculation in order to have the full picture of the market functioning during the chosen period. As a result we obtained two variants of the market behavior. DJI and S&P (fig. 9) reflect the first one. These markets contain the period of crisis events and relaxation after the Dotcom bubble burst, period of normal functioning, crisis and relaxation. All other markets refer to the second variant. The difference is that it contains the instability period before the crisis. Let's consider, for example, FTSE data series and its corresponding LAM measures (fig. 12).

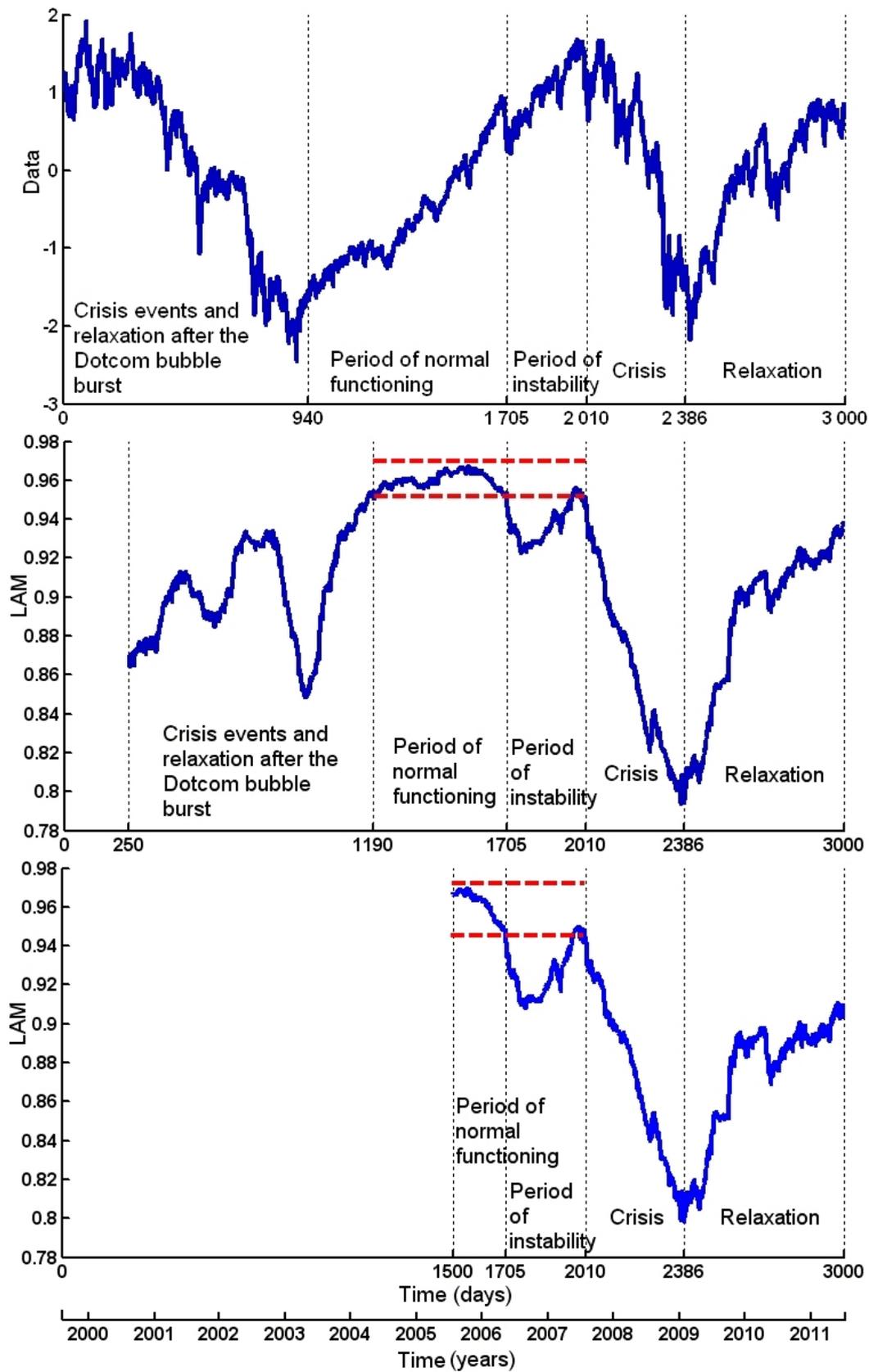

Fig. 12. FTSE from 23.08.1999 till 11.07.2011 and corresponding LAM.

According to the LAM measures, during the period from 20.12.2005 till 06.08.2007 (x = 1705 - 2010) FTSE experienced crisis events and recreation from them. It was not a real crisis, but the increasing of the market volatility because of the price rally, so it is better to name it the period of instability. For the majority of studied markets this period is approximately the same (table 1).

Table 1. Time characteristics of the crisis 2007 - 2010.

| Index | Period of normal functioning | Instability period | Crisis period | Relaxation beginning | Crisis time, days |
|---|---|---|---|---|---|
| DJI | 31.07.2003 - 27.08.2007 | - | 28.08.2007 - 22.04.2009 | 23.04.2009 | 416 |
| S&P | 10.03.2003 - 29.07.2007 | - | 30.07.2007 - 17.05.2009 | 18.05.2009 | 450 |
| FTSE | 15.05.2003 - 23.05.2006 | 24.05.2006 - 06.08.2007 | 07.08.2007 - 01.02.2009 | 02.02.2009 | 376 |
| DAX | 10.09.2003 - 24.05.2006 | 25.05.2006 - 18.11.2007 | 19.11.2007 - 07.05.2009 | 08.05.2009 | 370 |
| HSI | 16.07.2002 - 01.03.2006 | 02.03.2006 - 28.10.2007 | 29.10.2007 - 08.01.2009 | 09.01.2009 | 300 |
| Nikkei | 02.08.2004 - 12.02.2006 | 13.02.2006 - 02.08.2007 | 03.08.2007 - 26.01.2009 | 27.01.2009 | 360 |
| CAC | 11.07.2003 - 22.05.2006 | 23.05.2006 - 13.08.2007 | 14.08.2007 - 21.12.2008 | 22.12.2008 | 347 |
| ASE | 17.07.2001 - 30.05.2006 | 01.06.2006 - 01.05.2008 | 02.05.2008 - 02.12.2008 | 03.12.2008 | 150 |
| SMI | 10.06.2004 - 14.03.2006 | 15.03.2006 - 26.02.2007 | 27.02.2007 - 11.12.2008 | 12.12.2008 | 450 |
| OMXC | 20.09.2002 - 15.05.2006 | 16.05.2006 - 30.07.2007 | 31.07.2007 - 26.05.2009 | 27.05.2009 | 450 |
| PSI | 30.01.2004 - 28.07.2006 | 29.07.2006 - 29.07.2007 | 30.07.2007 - 03.12.2008 | 04.12.2008 | 345 |
| AORD | 08.09.1999 - 25.05.2006 | 26.05.2006 - 26.07.2007 | 27.07.2007 - 09.12.2008 | 10.12.2008 | 350 |

The stock prices rally is most likely to determine the period of instability during 2006 – 2007 years. Unfortunately, it is not obvious why the USA market didn't have this period. In any case RQA is able to reveal such differences in market behavior. Moreover, dates of the crisis and relaxation beginning coincide for those indexes that were analyzed in [Piskun & Piskun, 2011], in spite of the various length of rows – 3000 and 1500 points respectively. This shows, one more time, the stability of obtained results and noncriticality of the method to the time series length.

According to the table 1, the first indexes that recreated after the Dotcom bubble were AORD in 1999 and ASE in 2001. In 2002 fully relaxed HIS and OMXC. The period of normal functioning began in 2003 year for DJI, S&P, FTSE, DAX and CAC.

Later, in 2004 it started for Nikkei and SMI. Such divergent time dates indicate the various levels of involvement of different countries to the NASDAQ bubble 1999 - 2000. Economic interdependencies and the activity of investment in IT companies determined the difference in the recreation period.

DJI and S&P refer to the first variant of the market behavior during the crisis 2007 – 2010 and don't have the instability period. All other indexes began to experience it from 2006 year. For HSI, Nikkei and SMI this period started in the first quarter of 2006. FTSE, DAX, CAC, ASE, OMXC and AORD felt it in the middle of the year, and PSI in the third quarter.

The financial crisis started for SMI at the beginning of 2007. DJI, S&P, FTSE, Nikkei, CAC, OMXC, PSI and AORD began to experience it in the third quarter of 2007. In the case of DAX and HIS, crisis took place closer to the end of the year. The only one index that got it in 2008 was ASE.

Relaxation or recreation of the markets goes after the crisis and signalizes about the dumping of the negative economic tendencies and stabilization. Market agents adjusted to the new conditions and infrastructure began to rise up. The first indexes that got out from the crisis at the end of 2008 were CAC, ASE, SMI, PSI and AORD. At the beginning of 2009 relaxation started for FTSE, HIS and Nikkei. Closer to the middle of the year it began for DJI, S&P, DAX and OMXC.

The longest crisis period experienced S&P, SMI, OMXC – 450 and DJI – 416 financial days. For FTSE, DAX, Nikkei, CAC, PSI and AORD it lasted around $350 \pm 20$ days. HIS had crisis 300 days. For ASE this period was the shortest – 150 days.

The level of synchronization of the analyzed markets can be determined according to the instability and crisis period beginning. The time when relaxation starts or duration of the crisis can't be helpful as every country has their own methods of recreation, and coincidence of those dates wouldn't indicate the tough interrelations between countries. So, the difference in the instability occurrence on the markets, besides DJI and S&P, is limited by several months. This shows the involvement of all countries in the stock prices rally. The crisis period also began within some months. Only ASE and SMI are exceptions. We can come to the conclusion that the majority of countries with the developed stock market are highly synchronized. However, the further research to support this statement is needed.

After we have the time characteristics of the crisis 2007 - 2010, let's consider whether there are any quantitative regularities in LAM behavior (table 2). This time we used the same date series, but applied the new method of LAM calculation with LRP = 1500. This way, we obtained LAM graphs as in the case of everyday monitoring in real time.

Values of LAMs during the period of normal functioning of stock indexes are from 0,94 till 0,99. In [Bastos & Caiado, 2011] it was showed that LAM of emerging markets is higher than of the developed ones. Here we can't reveal any differences in markets as all of them are in one group. At the same time the values of LAM are slightly diverse, because of the inequality of volatility.

The LAM minimum indicates the last crisis point, after which, the relaxation begins. The drop of the LAM value from the point that corresponds to the crisis beginning to the point of its ending characterizes the power of crisis. For DJI, S&P and FTSE crisis was rather severe. They have the highest LAM drops. A little bit lower decreasing of LAM have HSI, Nikkei and CAC. DAX, SMI, OMXC and AORD suffered from crisis less than the previous indexes.

Table 2. Quantitative characteristic of the crisis 2007 - 2010.

| Index | LAM of the period of normal functioning | LAM minimum | LAM drop | LAM drop, % |
|---|---|---|---|---|
| DJI | 0,94 - 0,95 | 0,735 | 0,205 | 21,81% |
| S&P | 0,94 - 0,953 | 0,776 | 0,164 | 17,45% |
| FTSE | 0,945 - 0,97 | 0,7898 | 0,1802 | 18,58% |
| DAX | 0,98 - 0,985 | 0,885 | 0,095 | 9,69% |
| HSI | 0,965 - 0,967 | 0,84 | 0,125 | 12,95% |
| Nikkei | 0,97 - 0,978 | 0,851 | 0,119 | 12,27% |
| CAC | 0,963 - 0,982 | 0,856 | 0,107 | 11,11% |
| ASE | 0,97 - 0,984 | 0,905 | 0,065 | 6,70% |
| SMI | 0,955 - 0,975 | 0,873 | 0,082 | 8,59% |
| OMXC | 0,948 - 0,955 | 0,865 | 0,083 | 8,76% |
| PSI | 0,985 - 0,99 | 0,925 | 0,06 | 6,09% |
| AORD | 0,94 - 0,945 | 0,865 | 0,075 | 7,98% |

The lightest negative impact got ASE and PSI. So, we can reveal the level of crisis influence on the markets, but are not able to form any quantitative regularities of the LAM behavior. However the prognostication, in such case, is impossible, the visual analysis of the LAM graph with the help of an expert or indication of the control points (crisis beginning and ending) by means of automated system might give appropriate information about the market state.

The next step of the present study is the currency market. It plays a significant role in the world of finance. The stability of the exchange rate determines the effectiveness of the international business. That is why, basic six currency pairs (AUD/USD, EUR/USD, GBP/USD, USD/CAD, USD/CHF, USD/JPY) were taken for the analysis. FOREX differs from the stock market and longer time series for the research are required – 3500 points from 11.07.2011 and back to the past; LPR here equal 2000. Analogically to the analysis of stocks, we used the regular method of LAM calculation for the time characteristic of the crisis and the new one for the quantitative description. Let's consider GBP/USD pair from 06.08.01 till 11.07.2011 (fig. 13).

Presented currency pair has the period of crisis events and relaxation after the Dotcom bubble burst, period of normal functioning, period of instability, crisis and relaxation. The same periods have all other currencies. Time characteristics of them are introduced in table 3.

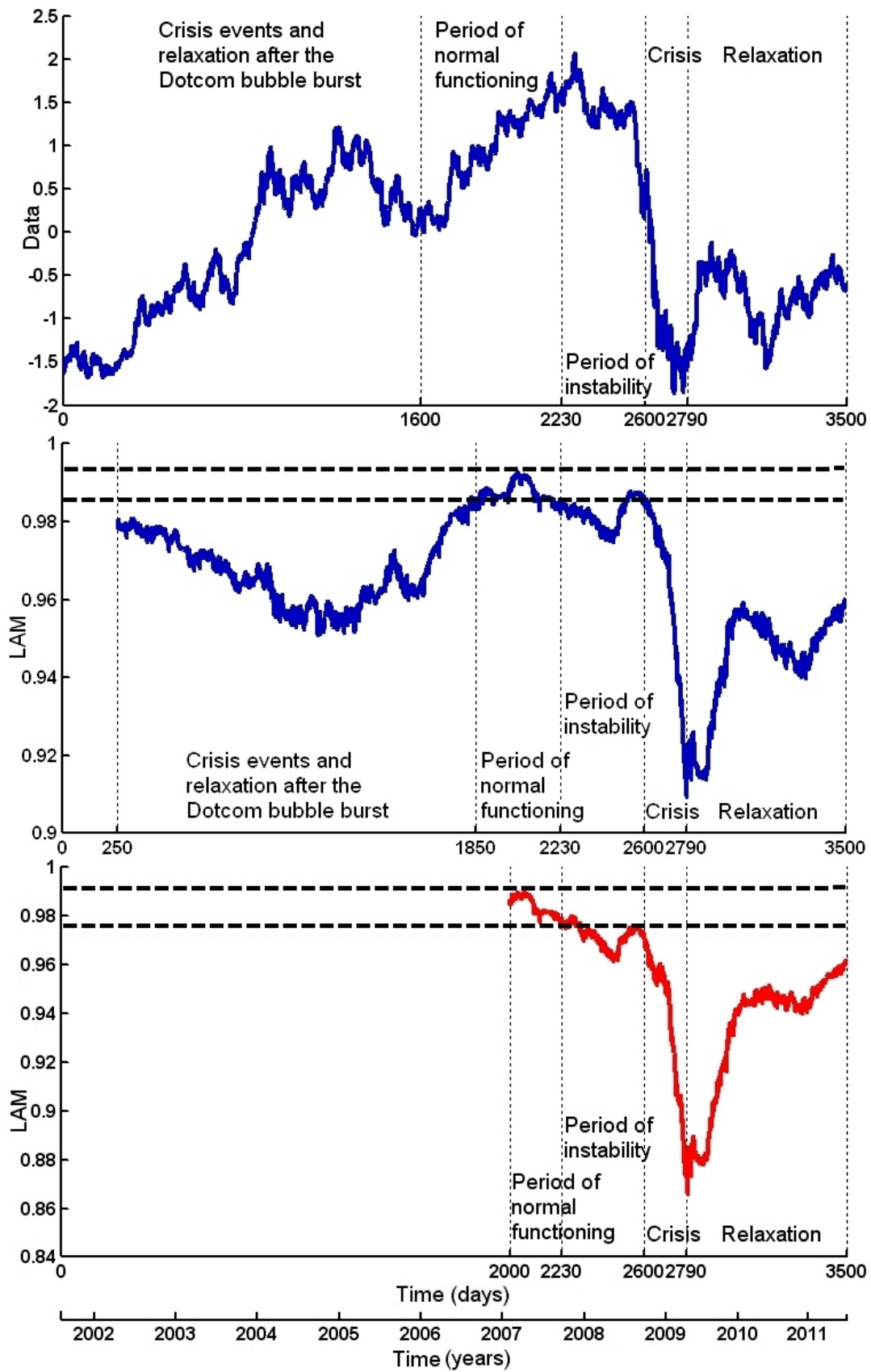

Fig. 13. GBP/USD from 06.08.21 till 11.07.2011 and corresponding LAM measures.

Table 3. Time characteristic of the crisis 2007 - 2010.

| Currency pair | Period of normal functioning | Instability period | Crisis period | Relaxation beginning | Crisis time, days |
|---|---|---|---|---|---|
| AUD/USD | 12.09.2005 - 13.08.2007 | 14.08.2007 - 27.07.2008 | 28.07.2008 - 02.06.2009 | 03.06.2009 | 295 |
| EUR/USD | 22.12.2005 - 20.04.2008 | 21.04.2008 - 30.11.2008 | 01.12.2008 - 09.06.2009 | 10.06.2009 | 175 |
| GBP/USD | 22.12.2005 - 13.09.2007 | 14.09.2007 - 16.09.2008 | 17.09.2008 - 31.03.2009 | 01.04.2009 | 190 |
| USD/CAD | 09.02.2006 - 22.09.2007 | 23.09.2007 - 04.11.2008 | 05.11.2008 - 20.06.2009 | 21.06.2009 | 210 |
| USD/CHF | 22.12.2005 - 11.03.2008 | 12.03.2008 - 15.12.2008 | 16.12.2008 - 06.08.2009 | 07.08.2009 | 210 |
| USD/JPY | 22.12.2005 - 03.10.2007 | 04.10.2007 - 28.07.2008 | 29.07.2008 - 09.06.2009 | 10.06.2009 | 300 |

Four currency pairs - EUR/USD, GBP/USD, USD/CHF and USD/JPY fully recreated from the Dotcom bubble burst at the same day. AUD/USD relaxed some months before them and USD/CAD – a couple of month later than they. The instability period began in the third and fourth quarters of 2007 year for AUD/USD, GBP/USD, USD/CAD and USD/JPY. Later, in 2008 it started for EUR/USD and USD/CHF. The crisis occurred in the second half of 2008. First pairs that experienced it were AUD/USD and USD/JPY. Then it touched GBP/USD, USD/CAD, EUR/USD and USD/CHF. Markets moved to the phase of relaxation in the middle of 2009 year. GBP/USD is the first pair that got out from the crisis (01.04.2009). AUD/USD, EUR/USD, USD/CAD and USD/JPY did it in the sixth month and USD/CHF in the eighth. The longest time of crisis lasting have AUD/USD and USD/JPY – 295 and 300 financial days accordingly. USD/CAD and USD/CHF suffered from it 210 days. The shortest crisis period is observed for EUR/USD and GBP/USD – less than 200 days.

According to the table 3 instability and crisis beginning dates differ between the currency pairs. We can't say that disparity is rather significant, but at the same time it reaches up to six months. On the other hand, general tendency of the market behavior is similar. So, speaking about the level of synchronization the additional study is required. Lets move to the quantitative description of the crisis (table 4).

Table 4. Quantitative characteristic of the crisis 2007 - 2010.

| Index | LAM of the period of normal functioning | LAM minimum | LAM drop | LAM drop, % |
|---|---|---|---|---|
| AUD/USD | 0,98 - 0,99 | 0,79 | 0,19 | 19,39% |
| EUR/USD | 0,98 - 0,99 | 0,845 | 0,135 | 13,78% |
| GBP/USD | 0,981 - 0,99 | 0,866 | 0,115 | 11,72% |
| USD/CAD | 0,987 - 0,995 | 0,86 | 0,127 | 12,87% |
| USD/CHF | 0,977 - 0,993 | 0,837 | 0,14 | 14,33% |
| USD/JPY | 0,96 - 0,97 | 0,825 | 0,135 | 14,06% |

The LAM values of the period of normal functioning are rather similar for the analyzed pairs. The exception is USD/JPY. The department of currency regulation in Japan works the whole day long without weekends and provides the policy of the cheap yen. Probably, because of this tough supervision, USD/JPY has the lower values of LAM.

It seems that the crisis was the most hard for AUD/USD, because it has the highest LAM decline. All other currency pairs have smaller drops of rates and differ in two percent, so they lighter experienced the crisis. In case of stock markets it would be a right conclusion, but for the currency ones it is not always true. Each pair contains the values of two different currencies. When the one value decreases and the other one is constant – the rate of the pair becomes more turbulent. Nevertheless, when both values fall – the market volatility would not increase significantly, because the exchange rate stays more or less stable. That is why it is hard to estimate what currency pair got the hardest crisis impact.

Any quantitative regularities of LAM behavior that are useful in crisis forecasting were not revealed. So, similarly to stocks, LAM can be used for the market state monitoring in real time.

International market of commodities is also worthwhile to be considered. The majority of investors put their money into metals when financial assets experience instability. Moreover, the prices on such resources, for example, like oil determine the effectiveness of the development of the private business and the whole countries as well. So, for our study oil, gasoline, gold and silver price series from 11.07.2011 and 3000 points to the past were taken. The parameter LPR was chosen 1500 points. Let's have a look on the oil graph with two corresponding LAMs (fig. 14).

According to the figure, behavior of the new LAM differs from the regular's one. Before the point of crisis beginning (x = 2085) the new LAM demonstrates the increasing trend instead of decreasing one. After the crisis starts, dynamics of two LAMs become similar. The same situation is observed for all analyzed commodities. Speaking about stocks and currencies, only DJI has such kind of the new LAM behavior (fig. 9). It is hardly to say why it is so, but it should be definitely taken into account during the development of the automated monitoring system.

Oil contains normal period of functioning from 10.08.1999 till 30.07.2004 (x = 0 - 1244). Instability began on the 31.08.2004 (x = 1245) and 06.12.2007 (x = 2085) the crisis started. The market moved to the relaxation period on the 28.01.2009 (x = 2373), but from 08.03.2010 (x = 2650) new crisis events took place. The analysis of other commodities presented in table 5.

First of all, let's pay attention to the fact that Dotcom bubble and NASDAQ crash of 2000 had no influence on this market. Analyzed commodities have also rather various dates of the instability beginning. For gasoline it's the third quarter of 2003, for gold it's the end of 2005 and for silver – the middle of 2006. The crisis beginning is not similar as well. Gasoline felt it in the first quarter of 2008, gold – third quarter of 2006 and silver – in the middle of 2010. The point of relaxation beginning occurred in the first month of 2009 for gasoline and in the middle of 2009 for gold. Silver has experienced crisis till present time. After some period of recreation, new crisis events came to the market. Gasoline got them in the second quarter of 2011 and gold – in the second quarter of 2010.

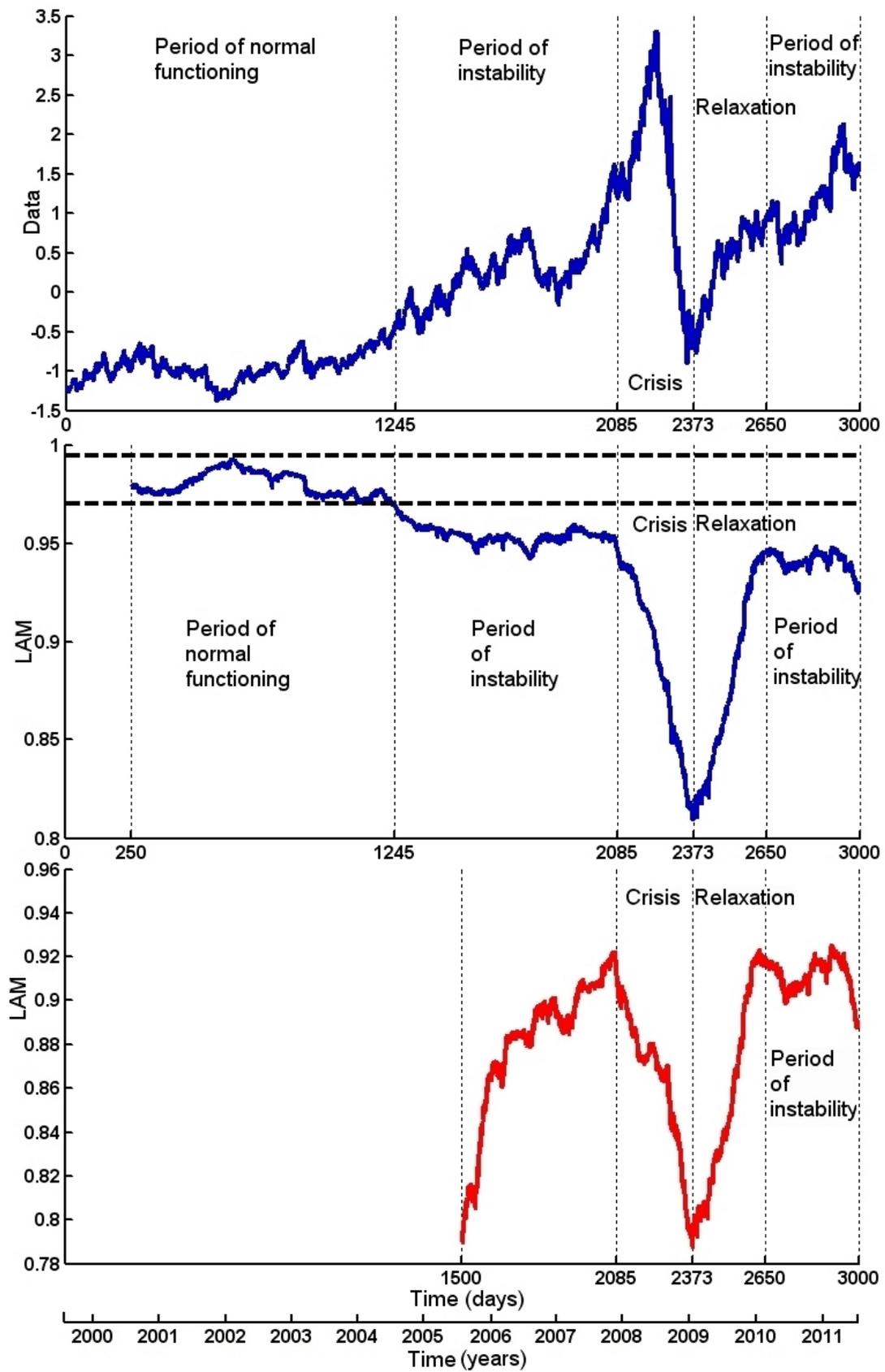

Fig. 14. Oil from 26.07.1999 till 11.07.2011 and corresponding LAM measures.

Table 5. Time characteristics of the crisis 2007 - 2010.

| Index | Period of normal functioning | Instability period | Crisis period | Relaxation beginning | Crisis events beginning | Crisis time, days |
|---|---|---|---|---|---|---|
| Oil | 10.08.1999-30.07.2004 | 31.07.2004-05.12.2007 | 06.12.2007-28.01.2009 | 29.01.2009 | 08.03.2010 | 287 |
| Gaso-line | 10.08.1999-08.09.2003 | 09.09.2003-11.03.2008 | 12.03.2008-16.01.2009 | 17.01.2009 | 13.05.2011 | 215 |
| Gold | 10.08.1999-22.12.2005 | 23.12.2005-04.10.2006 | 05.10.2006-18.06.2009 | 19.06.2009 | 13.04.2010 | 425 |
| Silver | 15.09.1999-22.06.2006 | 23.06.2006-07.06.2010 | 08.06.10-present time | - | - | - |

According to the significant differences in the assets behavior in comparison to each other and to those of the stock market, we can come to the conclusion of high autonomy of each single commodity. This market has its own laws of functioning and assumes the influence of external tendencies, such as stock market crash of 2000 or world financial crisis 2007 – 2010, in its own way.

Let's consider the quantitative characteristics of the crisis, obtained with the new LAM (table 6).

Table 6. Quantitative characteristic of the crisis 2007 - 2010.

| Index | Crisis beginning, LAM | LAM minimum | LAM drop | LAM drop, % |
|---|---|---|---|---|
| Oil | 0,915 | 0,787 | 0,128 | 13,99% |
| Gasoline | 0,860 | 0,774 | 0,086 | 10,00% |
| Gold | 0,945 | 0,882 | 0,063 | 6,67% |
| Silver | 0,926 | - | - | - |

Drops of the LAM differ essentially. Oil has 13,99 % decreasing, gasoline 10 % and gold – 6,67 %. For sliver the decline is not over jet. So, as in the cases of stock and currency markets, forecasting of the commodity crises is limited. Nevertheless, RQA is still useful for the effective state monitoring in real time.

## 5. Conclusions

In this work the assumption of the ability of laminarity measure of RQA to serve as a monitoring tool was checked. At the first stage, we determined appropriate embedding parameters (*m, τ, e* and *ws*) for the adequate RQA application and chose suitable measures for the financial time series analysis with the aim of the monitoring system development. Then, the process of LAM calculation in real time was simulated and tested on stock, currency and commodity markets. The study showed the measure's aptitude to distinguish the same periods of the market functioning as in the case of the static (regular) analysis. So, on the basis of laminarity, the monitoring system for the financial assets can be evolved.

Unfortunately, clear patterns of LAM behavior before and during the crisis period were not revealed, that leads to the impossibility of such events prediction.